\documentclass[aip,twocolumn,showpacs,floatfix,10pt,superscriptaddress,jcp]{revtex4-2}
\usepackage{graphicx}
\usepackage{color}
\usepackage{tikz}
\usepackage{epstopdf}

\newcommand{\be}{\begin{equation}}
\newcommand{\ee}{\end{equation}}
\newcommand{\bea}{\begin{eqnarray}}
\newcommand{\eea}{\end{eqnarray}}

\begin{document}

\title{Hard core lattice gas  with third next-nearest neighbor exclusion on  triangular lattice: one or two phase transitions? }
\author{Asweel Ahmed A. Jaleel}
\email{asweel@imsc.res.in}
\affiliation{The Institute of Mathematical Sciences, C.I.T. Campus,
Taramani, Chennai 600113, India}
\affiliation{Homi Bhabha National Institute, Training School Complex, Anushakti Nagar, Mumbai 400094, India}
\author{Dipanjan Mandal}
\email{dipanjan.mandal@warwick.ac.uk}
\affiliation{Department of Physics, University of Warwick, Coventry CV4 7AL, United Kingdom}
\author{R. Rajesh}
\email{rrajesh@imsc.res.in}
\affiliation{The Institute of Mathematical Sciences, C.I.T. Campus,
Taramani, Chennai 600113, India}
\affiliation{Homi Bhabha National Institute, Training School Complex, Anushakti Nagar, Mumbai 400094, India}

\date{\today}

\begin{abstract}
We obtain the phase diagram of the hard core lattice gas with third nearest neighbor exclusion on the triangular lattice using Monte Carlo simulations that are based on a rejection-free flat histogram algorithm. In a recent paper [J. Chem. Phys. {\bf 151}, 104702 (2019)], it was claimed that the lattice gas with third nearest neighbor exclusion undergoes two phase transitions with increasing density, with the phase at intermediate densities exhibiting hexatic order with continuously varying exponents. Though a hexatic phase is expected when the exclusion range is large, it has not been seen earlier in  hard core lattice gases with short range exclusion. In this paper, by numerically determining the entropies for all densities, we show unambiguously that there is only a single phase transition in the system between a low-density fluid phase and a high-density ordered sublattice phase, and that a hexatic phase is absent. The transition is shown to be first order in nature and the critical parameters are determined accurately.
\end{abstract}


\maketitle

\section{\label{sec:intro}Introduction}
Models of particles interacting through only excluded volume repulsion have been studied as minimal models for critical phenomena, self assembly, adsorption, etc. Well known examples studied on lattices include rods~\cite{2007-gd-epl-on,2013-krds-pre-nematic,2017-gkao-pre-isotropic,2017-vdr-jsm-different}, dimers~\cite{1961-k-physica-statistics,1961-tf-pm-dimer,2003-hkms-prl-coulomb}, squares~\cite{1967-bn-jcp-phase,1966-bn-prl-phase,1966-rc-jcp-phase,2012-rd-pre-high,2016-ndr-epl-stability,2016-nr-jsm-high,2017-mnr-jsm-estimating}, cubes~\cite{vigneshwar2019phase}, rectangles~\cite{2014-kr-pre-phase,2015-kr-pre-asymptotic,2015-nkr-jsp-high,2017-gvgmv-jcp-ordering}, triangles~\cite{1999-vn-prl-triangular}, tetrominoes~\cite{2002-mhs-jcp-simple,2009-bsg-langmuir-structure}, Y-shaped particles~\cite{szabelski2013selfassembly,2015-rthrg-tsf-impact,2018-pre-mnr-phase}, and hexagons~\cite{1980-b-jpa-exact}. 
Different shapes studied in the continuum include spheres~\cite{1957-aw-jcp-phase,1957-wj-jcp-preliminary, 2015-jcp-ik-hard}, polyhedra~\cite{2013-prl-ggrd-phase, 2012-jcp-mzl-freezing, 2010-pre-bst-phase}, plates~\cite{cuetos-2007-phase, rafael-2020-self}, etc. Despite having been well studied, it is still not possible to predict the different phases that exist for a given shape, and how the order of appearance depends on density.

Of special interest is the hard sphere model which is a minimal model for the freezing transition from fluid to solid. In three dimensions it undergoes a single first order phase transition from a fluid to a solid~\cite{hoover1968melting}. In two dimensional continuum, the system of hard disks freezes in a two-step process~\cite{1973-kt-jpc-ordering,1979-y-prb-melting,1979-nh-prb-dislocation}. As density is increased, the system first undergoes a transition from fluid to hexatic phase. The hexatic phase has power law orientational correlations with an exponent that changes with density. At higher densities, it undergoes a transition from the hexatic phase to a solid phase characterized by orientational order and power law correlations in positions. The liquid-hexatic phase transition is first order while the hexatic-solid phase transition is continuous~\cite{2011-bk-prl-twostep,engel2013hard,kapfer2015two}. 

The lattice model of hard spheres  is the $ k $-NN hard core lattice gas in which a particle excludes all the sites up to the $ k $-th next-nearest neighbors from being occupied by another particle. It is expected that as $ k $ increases, the discretization effects become insignificant and the continuum results will be recovered. The  $ k $-NN model has a long history, having been studied from the 1950s~\cite{1958-d-nc-theoretical,1960-b-pps-lattice,1961-b-pps-lattice,1966-bn-prl-phase}. We summarize what is known. On the square lattice (see~\cite{2007-fal-jcp-monte,nath2014multiple,2016-nr-jsm-high} and references within), models with $k=1, 2, 3, 5, 6, 7, 8, 9$ have been shown to undergo a single transition while $k=4, 10, 11$ undergo multiple phase transitions with increasing density. The multiple transitions do not include a hexatic phase and are due to the presence of a sliding instability at full packing~\cite{nath2014multiple,2016-nr-jsm-high} . On the honeycomb lattice  \cite{thewes2020phase,darjani2021glassy},  1-NN and 4-NN models show a single transition while 2-NN and 5-NN models show two transitions. Surprisingly, there is no transition found for 3-NN. On the triangular lattice the 1-NN model, also known as the hard hexagon model, is the only exclusion model that is exactly solvable~\cite{1980-b-jpa-exact,baxterBook}. The 1-NN and 2-NN models undergo a single phase transition~\cite{1968-OB-JCP-Traingularlattice,1980-b-jpa-exact,bartelt1984triangular,zhang2008monte,darjani2017extracting,akimenko2019tensor}. Preliminary study of the 4-NN and 5-NN model suggest a single phase transition and an absence of a hexatic phase~\cite{akimenko2019tensor}. However, for the 3-NN model, while some studies argue for a single first order phase transition, others claim the existence of a hexatic phase sandwiched between fluid and solid phases.  This is the only lattice model in which a hexatic phase has been reported, making it of particular interest and will be the focus of this paper. We now summarize the known quantitative results for the 3-NN model.

The 3-NN model on the triangular lattice was initially studied by Orban and Bellemans using matrix methods and series expansion~\cite{1968-OB-JCP-Traingularlattice}. It was shown that system undergoes a single first order phase transition from fluid to solid at  a critical reduced chemical potential $ \mu_c= 4.7\pm 0.2$. Recently, in a short note, using the tensor renormalization group method, it was shown that the system has a single first order phase transition at $\mu_c=4.4488 $~\cite{akimenko2019tensor}. Contrary to these results, based on the Monte Carlo simulations with adsorption and desorption, it has been  claimed that two phase transitions exist: first from a low density fluid phase to an intermediate density hexatic phase and second from a hexatic phase to a high density solid-like sublattice phase~\cite{darjani2019liquid}. Both transitions were argued to be first order in nature.  The liquid and hexatic phases co-exist between packing fraction $0.877$ and $0.915$ while the hexatic and solid coexistence starts at surface coverage $0.952$~\cite{darjani2019liquid}. 

To resolve the contradictory results for the 3-NN model on the  triangular lattice, as well as to study further the hexatic phase, if it exists,  we carry out a detailed study  for all packing fractions varying from $0$ to $1$. The numerical study of hard core lattice gases typically suffers from equilibration issues either when the excluded volume is large or when the densities are high. Algorithms that include cluster moves are able to overcome these issues. An example is the transfer matrix based strip cluster update algorithm (SCUA) that updates strips that span the lattice in one attempt~\cite{2012-krds-aipcp-monte,2013-krds-pre-nematic}. This algorithm has been very useful to obtain the phase diagram of the $k$-NN model on square~\cite{nath2014multiple} and honeycomb~\cite{thewes2020phase} lattices as well as many other shapes~\cite{2013-krds-pre-nematic,2015-rdd-prl-columnar,2017-vdr-jsm-different,2018-pre-mnr-phase,vigneshwar2019phase}. More recently, this algorithm has been combined with flat histogram methods, enabling the determination of density of states for all densities~\cite{jaleel2021rejection}. By applying this strip cluster Wang Landau (SCWL) algorithm to the 3-NN model, we  show that there is only one phase transition between a low density fluid phase and a high density sublattice phase, and there is no signature of a hexatic phase. The phase transition is shown to be first order in nature. Using the non-convexity properties of the entropy at the transition, we obtain precise estimates for the critical parameters. We show that the transition occurs at the reduced chemical potential  $4.4641(3)$  with the fluid and solid phases coexisting between densities  $0.8482(1)$ and  $0.9839(2)$. The critical pressure is determined to be 0.6397(1).

The remainder of the paper is organized as follows.  In Sec.~\ref{sec:model}, we describe the 3-NN model and the flat histogram algorithm that we use to simulate the model.  Section~\ref{sec:results} contains the results, where we show the existence of only one phase transition as well as show in multiple ways the first order nature of the transition. Section~\ref{sec:summary} contains a summary and discussion.
 
\section{\label{sec:model} Model and the Monte Carlo algorithm}

\begin{figure}
	\includegraphics[width=0.8\columnwidth]{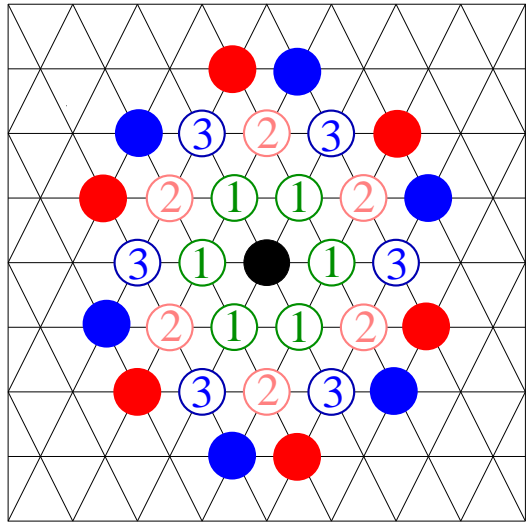}
	\caption{\label{fig:model} The first, second and third next-nearest neighbors of a particle (filled black circle) on the triangular lattice are denoted by 1, 2 and 3 respectively. In the 3-NN  model these sites are excluded from being occupied by another particle. The filled red and blue circles denote the fourth next-nearest neighbor sites.}
\end{figure}

Consider a two dimensional triangular lattice of linear size $L$ with periodic boundary conditions. A lattice site may be empty or occupied by utmost one particle. A particle excludes sites that are up to the third next-nearest neighbors (total of 18)  from being occupied by another particle (see Fig.~\ref{fig:model}). This model is referred to as the 3-NN model.   Since the interaction energies are either infinite or zero, temperature plays no role. We define reduced chemical potential, $\mu=\beta\tilde{\mu}$, $\tilde{\mu}$ is the chemical potential and  $ \beta =1/(k_B T)$, where $T$ is the temperature. Equivalently, we will work in units where $\beta=1$.

We  study this model mainly using SCWL algorithm~\cite{jaleel2021rejection} which combines the Wang Landau algorithm with rejection-free cluster moves.
The Wang Landau sampling algorithm is as follows~\cite{2001-wl-prl-efficient,2001-wl-pre-determining}. Let $S(N, L)$ denote the entropy of the system when $N$ particles are present. Then, 
\be 
S(N,L)=\ln g(N,L),
\ee
where $g(N,L)$ is the number of configurations with $N$ particles. The weight of a configuration with $N$ particles is taken to be proportional to $1/g(N, L)$.
Initially, $S(N,L)=0$  for all $N$.  Implement an evaporation-deposition algorithm (the details of evaporation and deposition are given below) that alters the number of particles consistent with the weights of configurations.  A histogram $H(N)$ records the number of times configurations with $N$ particles is visited. Every time a configuration with $N$ particles  is reached, $S(N, L) \to S(N, L)+f$, and $H(N) \to H(N)+1$. At the end of an iteration determined by the histogram being flat $(\min [H(N)] \geq c \max [H(N)])$ the factor $f$ is halved: $f \to f/2$, and the histogram is reset to zero. The initial value of $f$ is chosen to be $1$ and we chose $ c=0.80 $  . The above steps are repeated until $f$ is smaller than some pre-determined small value. In our simulation, we do 21 iterations, \textit{i.e.} the final value of $ f $ is $ 2^{-21} $.

We now describe the evaporation-deposition moves that we implement. In every move, we will generate a new configuration from many possible configurations based on their weights, so that rejections are avoided and at the same time low entropy configurations are accessed efficiently. We describe the steps below.

Consider an allowed configuration of 3-NN particles. Choose a row at random from the $3 L$ rows  (rows could be in any of the three lattice directions). The row breaks up into segments that are separated from each other by sites that are excluded from occupation due to the presence of particles in nearby rows.  From the identified segments, choose one at random.  Evaporate all the particles from this segment. Let the number of empty sites in the segment be $\ell$. After evaporation, let the number of particles present be  $N_0$. We will now refill this segment with a new configuration. Note that in this one dimensional segment, there  have to be at least two empty sites between two particles (see Fig.~\ref{fig:model}). Given $\ell$ sites, it is possible to put $0, 1, \ldots, [(\ell+2)/3]$ particles. We first determine the number of particles, $n$, that  should be deposited. Then, from the many configurations with $n$ particles, choose one at random. Once all the segments in the row are updated, we update the histogram and entropy.

We now address two questions: first, how do we decide the value of $n$? Second, given $n$, how do we choose a random configuration from the many possible configurations?

First consider a segment of length $\ell$ with open boundary conditions. 
Let $C_o(\ell,n)$ be the number of ways of occupying a  segment of length $\ell$ with $n$ particles. Then the probability ${\rm Prob}_o(\ell,n)$ of choosing $n$ particles to deposit is
\begin{equation}
{\rm Prob}_o(\ell,n)=\frac{C_o(\ell,n)/g(N_0+n,L)}{\sum_{i=0}^{n^*} C_o(\ell,i)/g(N_0+i,L)},
 \label{eqn:prbo}
\end{equation}
where $n^*=[(\ell+2)/3]$ is the maximum value $n$ can take.
Likewise, if the boundary conditions for the segment is periodic, then the probability ${\rm Prob}_p(\ell,n)$ of choosing $n$ particles to deposit is
\begin{equation}
{\rm Prob}_p(\ell,n)=\frac{C_p(\ell,n)/g(N_0+n,L)}{\sum_{i=0}^{n^*} C_p(\ell,i)/g(N_0+i,L)},
 \label{eqn:prbp}
\end{equation}
where $C_p(\ell,n)$ is the number of ways of placing $n$ particles on a ring of $\ell$ sites.

Determining $C_o(\ell,n)$ and $C_p(\ell,n)$ is a straightforward enumeration problem. Note that there must be at least two vacant sites between two neighboring particles. We obtain
\be
C_o(\ell,n)= \frac{(\ell+2-2n)!}{(\ell+2-3 n)! n!},~n=0,1,\ldots, \left[\frac{\ell+2}{3} \right]. \label{eqn:co}
\ee
$C_p(\ell,n)$ can be obtained from $C_o(\ell,n)$  through the   recursion relation:
\be
C_p(\ell,n) = 2 C_o(\ell-5,n-1) +C_o(\ell-2,n),
\label{eq:periodic-recursion}
\ee
where the first term on the right hand side describes putting a particle in one of the first two sites, and the second term describes the case when the first two sites are empty. From Eqs.~(\ref{eqn:co}) and (\ref{eq:periodic-recursion}), we obtain
\be
C_p(\ell,n)= \frac{\ell(\ell-2 n-1)!}{(\ell-3 n)! n!},~n=0,1,\ldots, \left[\frac{\ell}{3} \right]. \label{eqn:cp}
\ee

After determining $n$, we fill the segment iteratively from one end to other. For open segments, the probability $P_o(\ell,n)$ that the first site is empty is given by
\be
P_o(\ell,n)= \frac{C_o(\ell-1,n)}{C_o(\ell,n)}=\frac{\ell+2-3 n}{\ell+2-2n}.
\ee
If the first site is empty, then $\ell$ is reduced by one, keeping $n$ the same, and the step is repeated. If the first site is occupied, then $\ell \to \ell-2$, $n\to n-1$, and the step is repeated.
If the segment has periodic boundary conditions, the probability $P_p(\ell,n)$ that the first two sites are empty is given by
\be
P_p(\ell,n)= \frac{C_o(\ell-2,n)}{C_p(\ell,n)}=\frac{\ell-2n}{\ell}.
\ee
If the first two sites are empty, then it reduces to an open segment of length $\ell-2$ with $n$ particles. Else, if one of the first two sites is occupied, then it reduces to an open segment of length $\ell-5$ with $n-1$ particles.

$C_o(\ell,n)$ and $ C_p(\ell,n) $  do not change during the  simulation and are therefore  evaluated once in the beginning and stored as a look up table. All the readings of the thermodynamic quantities like order parameter, susceptibility, etc., are computed only in the final iteration. The error is computed from 16 independent simulations for each system size $ L $.


To bench mark the implementation of the flat histogram algorithm, we compare the results of our simulations with results from fixed fugacity grand canonical simulations using SCUA. The SCUA is described in Appendix~\ref{appa}. For $L=35$, the variation of the density and order parameter obtained from both methods of simulations are compared in Appendix~\ref{appb}. The data match very well.

	In the paper, we have used three different algorithms to generate the data for the figures. To improve readability, we summarize in Table~\ref{tab:methods}, the Monte Carlo method used to generate the data for each figure.

\begin{table}		
	\caption { The Monte Carlo methods used to generate the  data for each figure. F denotes flat histogram simulations, GC denotes fixed $\mu$ grand canonical simulations and C denotes fixed density canonical simulations. }
	\begin{ruledtabular}
		\begin{tabular}{|c|p{1.4cm}|p{1.4cm}|p{4.5cm}|}
			\hline
			Fig.& Algorithm &Ensemble & Goal   \\
			\hline
			2 & SCWL  & F & Variation of entropy \\				\hline
			3 &  SCWL& F & Variation of average density and compressibility \\			\hline
			5 &  SCUA & GC & Sublattice densities in fluid phase \\			\hline
			6& SCUA & GC & Sublattice densities in solid phase \\			\hline
			7&  Fixed density & C & Coexistence of phases\\			\hline
			8&  SCWL &F & Absence of second transition.\\	 		\hline
			9&  SCWL & F & First order nature\\			\hline
			10&  SCWL & F & First order nature\\			\hline
			11 & SCWL & F & Precise estimation of coexisting densities and chemical potential. \\	\hline	
			12 &   SCWL & F & Finite size scaling \\				\hline
			13 & SCWL & F & Finite size scaling\\			\hline
			14&SCUA&GC& Ability to equilibrate  for densities as large as $0.99$\\ \hline  
			15&SCUA and SCWL&GC and F& Comparison of algorithms\\ \hline  
		\end{tabular}
	\end{ruledtabular}
	\label{tab:methods}
\end{table}

\section{\label{sec:results}Results}

\begin{figure}
\includegraphics[width=\columnwidth]{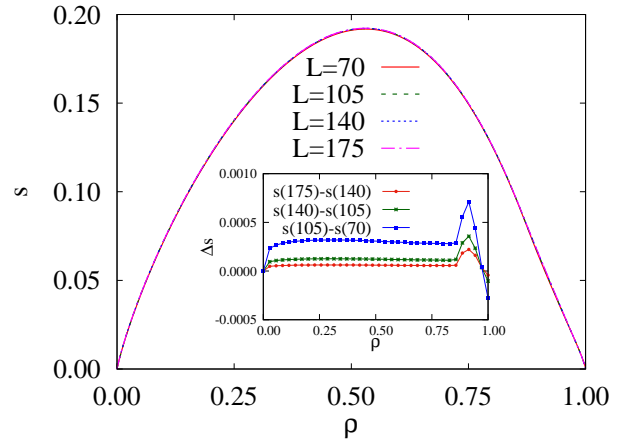}
\caption{\label{fig:entropy}  Variation of the entropy per site $ s $ with density $\rho$ for different system sizes. The entropy is normalized by setting $ S(0,L)=0$. Inset shows difference in $ s $ between two successive system sizes. The data are obtained using SCWL algorithm}
\end{figure}

We determine the entropy of the 3-NN model for system sizes up to $L=175$  using the flat histogram algorithm. Let density $\rho=\eta/\eta_{max}$, where $\eta$ is the number density and $\eta_{max}=1/7$ is the maximum number density.  In Fig.~\ref{fig:entropy}, the variation of the entropy per site, $s= S/L^2$ with $\rho$  is shown for different  $L$, where the entropy is normalized by setting $S(0,L)=0$.  We observe that the algorithm is able to easily access the fully packed state ($\rho=1$). Second, we see that beyond $L=70$, there is very little finite size effect, and the curves lie on top of each other. To see the convergence, the difference in entropies ($\Delta s$) between two successive $L$'s  are plotted in the inset of Fig.~\ref{fig:entropy}. The difference were calculated at a density interval of $\Delta \rho = 1/35$. The difference in the entropies  decreases with increasing $ L $.

Knowing entropy, the grand canonical partition function $\mathcal{L}(\mu,L)$ and pressure $ P $ are given by 
 \bea
 \mathcal{L}(\mu,L) &=&\sum_{N=0}^{N_{max}} g(N,L) e^{\mu N}, \label{eqn:pf} \\
P(\mu) &=&	\frac{1}{L^2} \ln	\mathcal{L}(\mu,L). \label{eqn:prm}
  \eea
The average density $\langle\rho\rangle$ and compressibility $\kappa$ is then given by 
\bea
\langle \rho \rangle &=& \frac{7}{\mathcal{L}(\mu,L)} \sum_{N=0}^{N_{max}} N g(N,L)  \mathrm{e}^{\mu N}, \label{eqn:avgr} \\
\kappa&=&L^2\left[ \langle \rho^2 \rangle -\langle \rho \rangle^2 \right]\label{eq:compressibility}, 
\eea

\begin{figure}
 \includegraphics[width=\columnwidth]{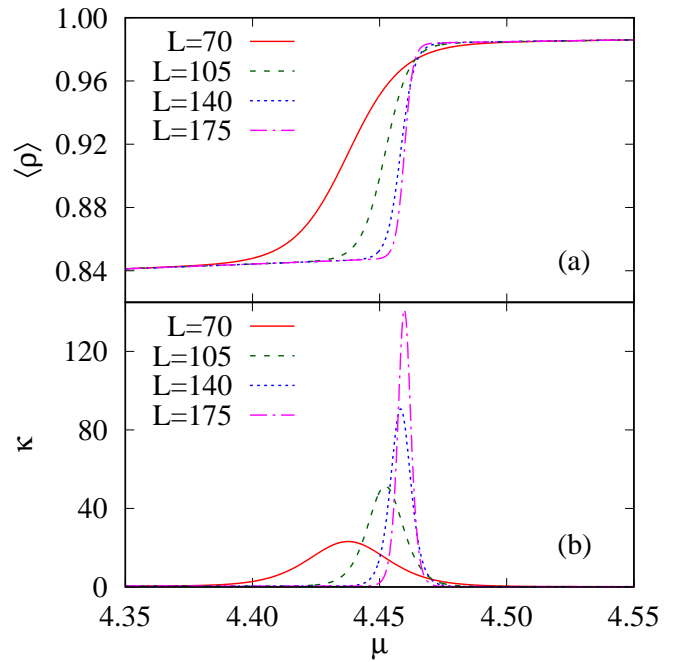}
 \caption{\label{fig:avgrk}  The variation of (a) average density $\langle\rho\rangle$ and  (b) compressibility $\kappa$ as a function of with chemical potential $\mu$.  The data are obtained using SCWL algorithm}
 \end{figure}

We first show that the system undergoes only one phase transition as the density is increased from 0 to $1$. A phase transition corresponds to a singular behavior in both density $\rho$ and compressibility $\kappa$. In particular, $\kappa$ typically diverges, though this need not always be the case (for example when the critical exponent, $\alpha <0$).  The variation of $\langle\rho\rangle$ and  $\kappa$ with chemical potential $\mu$ is shown in Fig.~\ref{fig:avgrk} for different $L$. The density $\rho$ has a discontinuity around $\mu_c \approx 4.46$ (see Fig.~\ref{fig:avgrk}(a)) while $\kappa$ diverges with system size at the same value of $\mu$ (see Fig.~\ref{fig:avgrk}(b)). Thus, there is at least one phase transition. 

We now argue that this is the only phase transition. To do so, we show that the phase for $\mu$ slightly smaller than $\mu_c$ is the disordered phase seen at low densities and  the phase for $\mu$ slightly larger than $\mu_c$ is the solid-like  sublattice phase seen at full packing.

\begin{figure}
 \includegraphics[width=\columnwidth]{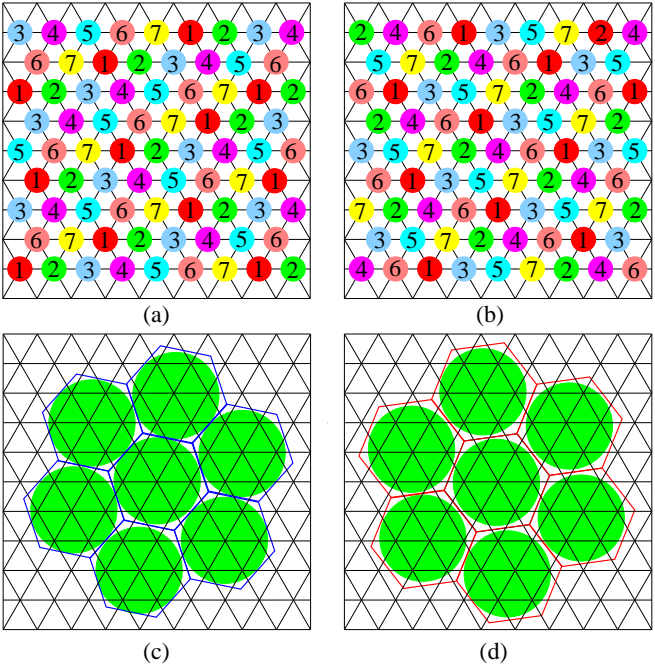}
 \caption{\label{fig:sublattice-shape} Panels (a) and (b) show two equivalent ways of dividing  the triangular lattice in to 7 sublattices. Panels (c) and (d) show the  effective shapes of particles at full packing corresponding to (a) and (b) respectively.}
 \end{figure}

To characterize the two phases, we divide the lattice into 7 sublattices based on the allowed configurations at full packing. In Fig.~\ref{fig:model}, the red and blue sites are the  fourth next-nearest neighbors. At full packing, either the blue or  the red sites are completely filled. This leads to two possible ways to divide the lattice into $7$ sublattices, which we denote as type-A and type-B, as shown in Fig.~\ref{fig:sublattice-shape}(a) and \ref{fig:sublattice-shape}(b) respectively. 
The effective space filling shape of particles for type-$A$ and type-$B$ sublattice divisions are shown in Fig.~\ref{fig:sublattice-shape}(c) and \ref{fig:sublattice-shape}(d) respectively.  In the disordered phase all the other 14 sublattices (7 of type-$A$ and 7 of type-$B$) will be on an average occupied equally  and in the sublattice phase, one of the 14 sublattices will be predominantly occupied.

\begin{figure}
\includegraphics[width=\columnwidth]{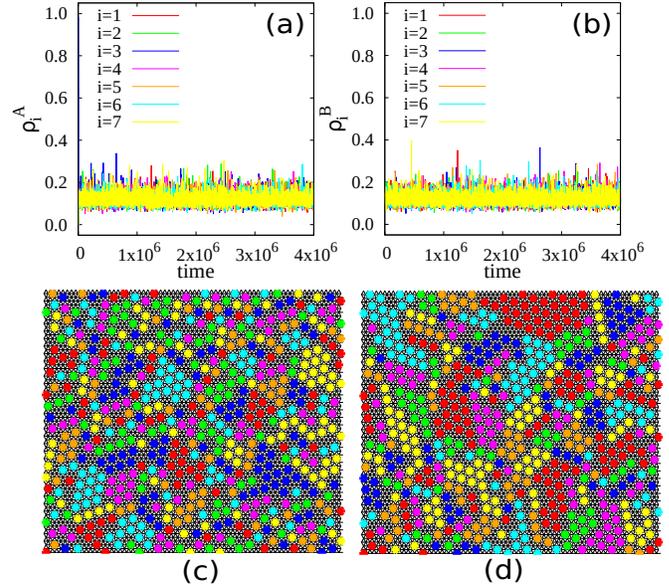}
\caption{The variation of sublattice densities with time and snapshots of typical equilibrated configurations  are shown for $\mu=4.38\lesssim \mu_c$ ($\langle\rho\rangle\approx 0.844$). The data are for $L=70$. $ \rho_i^X $ denotes the sublattice density of  $i$-th sublattice of type-X. In the snapshots, a particle is colored according to the sublattice it belongs to. Panels (c) and (d) correspond to type-A and type-B sublattice divisions respectively. The data are obtained using SCUA algorithm. }
\label{fig:snap_disordered}
\end{figure}

To confirm there is only one transition, we do fixed $\mu$ grand canonical simulations using SCUA (see Appendix~\ref{appa} for details about this algorithm) for $\mu=4.38\lesssim\mu_c$ and $\mu=4.50\gtrsim\mu_c$. The data for $\mu=4.38$  ($\langle\rho\rangle\approx 0.844$) are shown in  Fig.~\ref{fig:snap_disordered}. All the 14 sublattice densities are on an average equal [see Fig.~\ref{fig:snap_disordered}(a) and (b)]. Snapshots of typical equilibrated configurations, where the different sublattices are colored differently, show all colors with small domains of each sublattice [see Fig.~\ref{fig:snap_disordered}(c) and (d)]. We conclude that $\mu=4.38$ corresponds to the disordered fluid phase.

\begin{figure}
\includegraphics[width=\columnwidth]{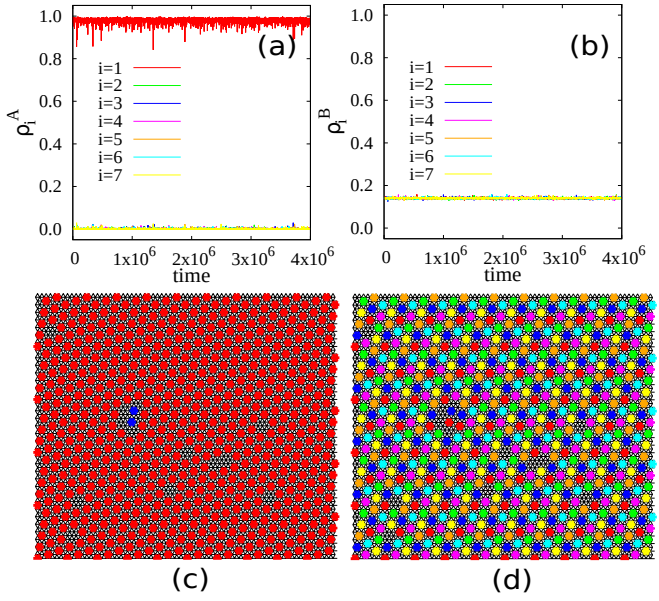}
\caption{The variation of sublattice densities with time and snapshots of typical equilibrated configurations  are shown for $\mu=4.50\gtrsim \mu_c$ ($\langle\rho\rangle\approx 0.985$). The data are for $L=70$. $ \rho_i^X $ denotes the sublattice density of  $i$-th sublattice of type-X. In the snapshots, a particle is colored according to the sublattice it belongs to. Panels (c) and (d) correspond to type-A and type-B sublattice divisions respectively. The data are obtained using SCUA algorithm.}

\label{fig:snap_sublattice}
\end{figure}

The data for $\mu=4.50$ ($\langle\rho\rangle\approx 0.985$) are shown in Fig.~\ref{fig:snap_sublattice}. One of the sublattice densities (in this instance a type-A sublattice) is larger than the others [see Fig.~\ref{fig:snap_sublattice}(a) and (b)]. The corresponding snapshots show predominantly one color for type-A sublattices and all colors for type-B sublattices [see Fig.~\ref{fig:snap_sublattice}(c) and (d)]. We conclude that $\mu=4.50$ corresponds to the solid-like sublattice phase. 

\begin{figure}
\includegraphics[width=\columnwidth]{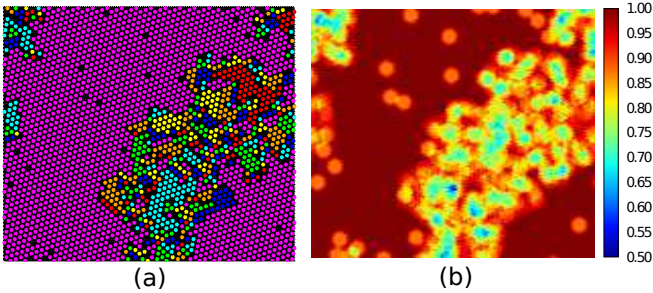}
\caption{(a) Snapshot of a typical configuration and (b) the corresponding density map at $\rho=0.920$, obtained from fixed-density simulations.  The system size is $L=140$.  The color scheme for the snapshot is same as that used in Figs.~\ref{fig:snap_disordered} and \ref{fig:snap_sublattice} . 
}
\label{fig:snap_canonical}
\end{figure}

We now check for the phase at intermediate densities by doing fixed-density Monte Carlo simulations (details of the algorithm are given in Appendix~\ref{appc}) at $\rho=0.92$. The snapshot of a typical equilibrated configuration and corresponding density map are shown in Fig.~\ref{fig:snap_canonical}(a) and (b) respectively. The coarse-grained density at a site is obtained by averaging the density over the sites up to the 7th next-nearest neighbor. From the snapshot we can clearly see co-existence of both sublattice phase and disordered state as the system has phase separated into a sublattice phase dominated by one color and a disordered phase with all 7 colors. From the density map, we can see the disordered phase has lower density. The observed coexistence of two phases rules out any possibility for an intermediate hexatic phase, and we conclude that there is only one phase transition.

\begin{figure}
		\includegraphics[width=\columnwidth]{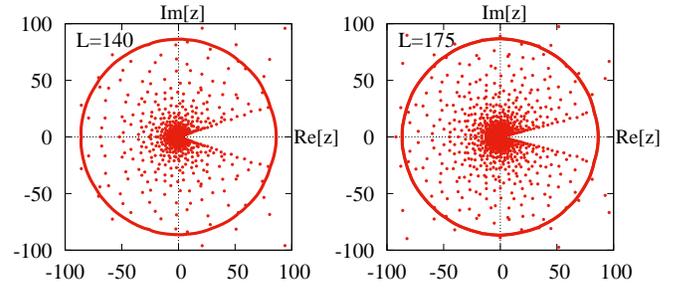}
		\caption{\label{fig:pfz} Zeros of the grand canonical partition function in the complex $ z $-plane ($z=\mathrm{e}^{\mu}$) for $  L=140 $ and $ L=175 $. The locus of the zeros form a circle and pinch the  positive real axis only one point. The data are obtained using SCWL algorithm.}
	\end{figure}

 Further strong evidence for only a single phase transition is obtained from the locus of the  zeros of grand canonical partition function [see Eq.~(\ref{eqn:pf})]~\cite{yang1952statistical,lee1952statistical}. Figure~\ref{fig:pfz} shows the zeros of the partition function for $L=140$ and $L=175$. The zeros pinch the positive $ z $-axis, where $z=\mathrm{e}^{\mu}$, at only one point and the locus is a circle. Since any phase transition corresponds to zeros approaching the real axis, the distribution of zeros in Fig.~\ref{fig:pfz} imply that there is only one phase transition. Also the locus being a circle implies a first order transition~\cite{creswick1997finite,taylor2013partition}.  In a  continuous transition, locus of zeros approaches the positive real axis at angle less than $\pi/2$~\cite{creswick1997finite,bena2005statistical,taylor2013partition}.

We now show that the transition is discontinuous and determine the critical parameters accurately. We first define an order parameter. Let
\bea
Q_A &=&\sum_{k=1}^{7} \rho_k^A\exp \left[\frac{2\pi i (k-1)}{7}\right],\\
Q_B &=&\sum_{k=1}^{7} \rho_k^B\exp \left[\frac{2\pi i (k-1)}{7}\right],
\eea
where $\rho_k^X$ is the $k$-th sublattice density for X-type  sublattice division.
Non-zero $Q_A$ or $Q_B$ implies that a particular sublattice is occupied preferentially. The order parameter $Q$ is defined to be
 \be
Q=||Q_A|-|Q_B||. \label{eqn:op}
\ee
In the disordered phase, all $14$ sublattice densities are on an average  equal and hence $Q=0$. In the sublattice phase one of $Q_A$ or $Q_B$ becomes non-zero, and hence $Q\neq 0$. 
The susceptibility $\chi$ and the Binder cumulant $U$ associated with  $Q$ are then defined as
\bea
\chi&=&L^2\big[\langle Q^2\rangle-\langle Q\rangle^2\big]\label{eq:susceptibility},\\
U&=&1-\frac{\langle Q^4\rangle}{3\langle Q^2\rangle^2}\label{eq:binder}.
\eea

\begin{figure}
	\includegraphics[width=\columnwidth]{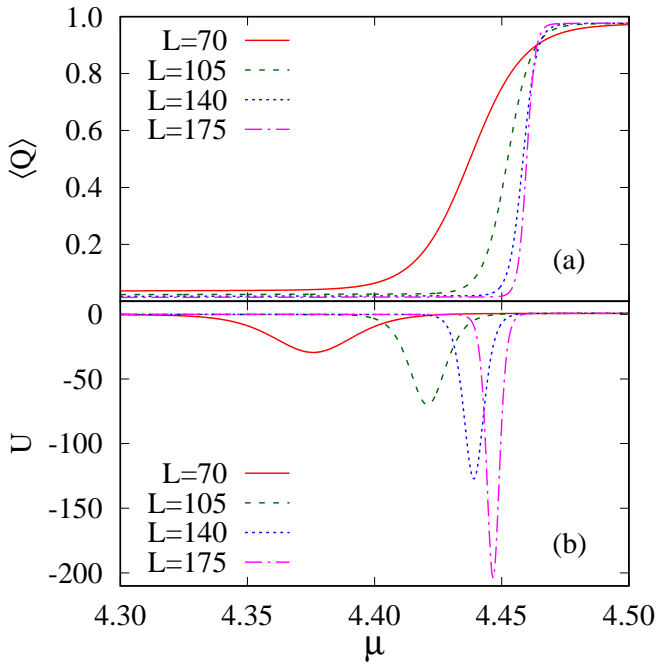}
	\caption{\label{fig:qu} Variation of the (a) order parameter $\langle Q \rangle$ and (b) Binder cumulant $U$ with chemical potential $\mu$.  The data are obtained using SCWL algorithm.}
\end{figure}

The variation of $Q$ and $U$ with $\mu$ is shown in Fig.~\ref{fig:qu} for different system sizes. $Q$  increases from $0$ to $1$ with increasing $\mu$, with the appearance of a discontinuity that becomes sharper with increasing system size, a signature of a first order transition. The Binder cumulant $U$ has a negative peak that increases with the system size $L$ which is a clear signature of  a first order phase transition~\cite{binder1984finite,vollmayr1993finite}.

More evidence for a discontinuous transition can be obtained from pressure. In the grand canonical ensemble, pressure is constant across a first order transition, while in the canonical ensemble, if homogeneity is assumed, pressure shows a non-monotonic behavior with density. The latter is usually corrected using the Maxwell construction. The grand canonical pressure $P$ is given in Eq.~(\ref{eqn:prm}). Let the canonical pressure be denoted by $\widetilde{P}$. Then,
\begin{equation}
\widetilde{P}=\int_{0}^{\rho} (1-\beta(\rho)) \frac{\partial}{\partial \rho} \left[ \frac{\rho}{1-\beta(\rho)}\right] d \rho, \label{eqn:prb}
\end{equation}  
where $1-\beta(\rho)$ is the mean fraction of sites where a new particle  can be occupied at density $\rho$~\cite{aveyard1973introduction,darjani2019liquid}. We compute $\beta(\rho)$ in the flat histogram simulations, allowing us to determine $\widetilde{P}$ from Eq.~(\ref{eqn:prb}).

\begin{figure}
	\includegraphics[width=\columnwidth]{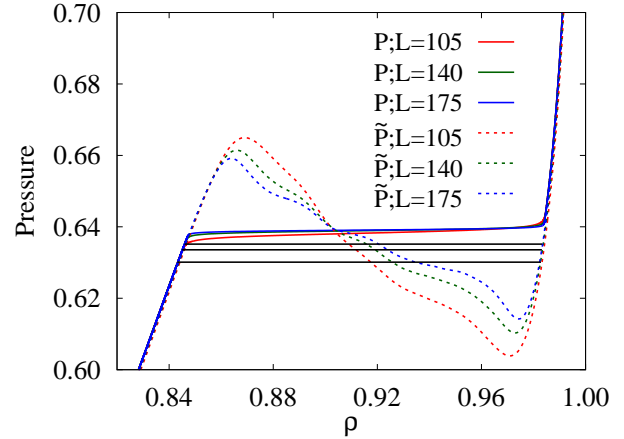}
	\caption{ Variation of the pressure with density $\rho$ for three different system sizes. The grand canonical pressure $P$ is calculated from Eq.~(\ref{eqn:prm}) and the canonical pressure $\widetilde{P}$ from Eq.~(\ref{eqn:prb}). The constant pressure lines shown in black solid lines  are obtained from the Maxwell equal area construction to $\widetilde{P}$ with the bottom line corresponding to $L=105$ and top line corresponding to $L=175$. As  $L$ increases from 105 to 175, the constant pressure lines move closer to the grand canonical $ P $. }
	\label{fig:pressure_beta}
\end{figure}

Figure~\ref{fig:pressure_beta} shows $P$ and $\widetilde{P}$ for three different system sizes. $P$ is near constant at the transition, while $\widetilde{P}$ shows non-monotonic behavior. By identifying the regions where $P$ and $\widetilde{P}$ differ, we identify approximately the co-existence densities $\rho_f$ and $\rho_s$ of the fluid and solid phases to be $\rho_{f}\approx 0.846$ and $\rho_{s}\approx 0.985$ for $L=105$.  We have also shown the constant pressure lines that are obtained from the Maxwell equal area construction to $\widetilde{P}$. As $ L $ increases, the constant pressure lines move closer to the grand canonical $ P $.

\begin{figure}
\includegraphics[width=\columnwidth]{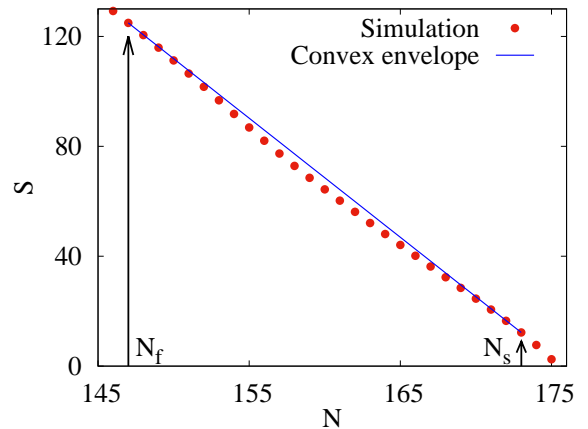}
\caption{ The non-convex part of the entropy and the corresponding convex envelope construction. $ N_f  $ and $ N_s $ denote the endpoints of the envelop in the fluid and solid phases respectively. The data are for $L=35$ obtained using SCWL algorithm.}
\label{fig:nconcave}
\end{figure}

The nature of the phase transition having been established, we now determine the critical chemical potential $\mu_c$, and the coexistence densities $\rho_f$ and $\rho_s$ more accurately.   To find  $\mu_c$, we use two methods: one based on convexity properties of entropy and the other based on susceptibility, $\chi$. The non-monotonicity in canonical pressure $\widetilde{P}$  with $\mu$ immediately suggests that, the measured entropy must be non-convex in a region of density. An example is shown in Fig.~\ref{fig:nconcave} for $L=35$.  The true entropy must be convex everywhere.  The pressure loop in the coexistence window of $\widetilde{P}$ in Fig.~\ref{fig:pressure_beta} is a finite size effect caused by the curved interface between a bubble of minority phase and the surrounding majority phase~\cite{2011-bk-prl-twostep}. This leads to the  non-convex behavior in entropy curve in the co-existence regime.

\begin{figure}
	\includegraphics[width=\columnwidth]{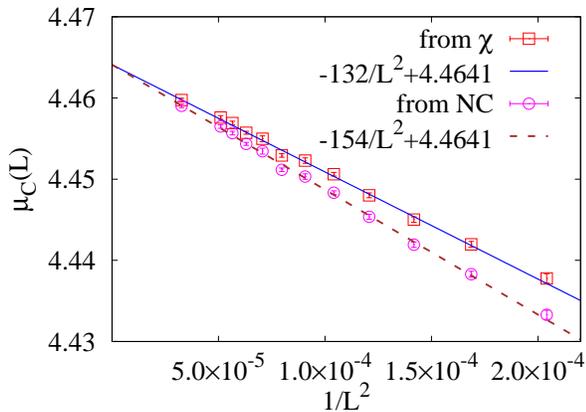}
	\caption{ Variation of critical chemical potential, $ \mu_c(L) $ with system size $ L $. The data are obtained from susceptibility, $\chi$ and non convexity (NC) analysis. Error in each data point was obtained from 16 independent simulations. Solid and dashed lines are the best linear fit to the data.}
	\label{fig:mubylsq}
\end{figure}

The slope of the convex envelope construction gives critical chemical potential, $\mu_c$:
\begin{equation}
	\mu_c(L)=-\frac{S(N_s)-S(N_f)}{N_s-N_f}.  \label{eqn:NC}
\end{equation}
$N_{f}$ and $N_{s}$ are number of particles at the boundaries of  the convex envelope and corresponds to  coexistence densities, $\rho_{f}$ and $\rho_{s}$ respectively. Position of  $N_{f}$ and $N_{s}$ are marked on the Fig.~\ref{fig:nconcave}. The system-size dependent critical parameters, thus obtained, are tabulated in Table~\ref{tab:2nn3d}. At a first order phase transition, the finite size corrections to the critical parameters decrease to zero as $L^{-2}$, i.e.,
\be
   \mu_{c} - \mu_{c}(L)  \sim  \frac{1}{L^2}. \label{eq:muscaling} 
\ee
Extrapolating to infinite $L$ (see Fig.~\ref{fig:mubylsq}), we obtain   $\mu_c=4.4641(3)$. The coexistence densities  also obey finite size scaling as in Eq.~(\ref{eq:muscaling}) and we obtain $\rho_f=0.8482(1)$ and $\rho_s=0.9839(2)$. 
	\begin{table}		
		\caption{Critical parameters from non convexity in entropy - 3NN. Extrapolation was performed against $1/L^2$ using linear regression. Error in each data point was obtained from 16 independent simulations. Error quoted in last row are from linear fit error.}
		\begin{ruledtabular}
		\begin{tabular}{|c|c|c|c|}
			\hline
			L& $\rho_{f}$ &$\rho_{s}$ & $ \mu_c $   \\
			\hline
70 &   0.84580(8) & 0.98571(9) & 4.43328(6) \\				\hline
77 &  0.84651(7)  & 0.98465(7) & 4.43829(4) \\				\hline
84 &  0.84654(11) & 0.98511(6) & 4.44192(3) \\			\hline
91 &  0.84694(5) & 0.98478(5) & 4.44536(3) \\			\hline
98 &  0.84693(5) & 0.98469(5) & 4.44832(2) \\			\hline
105&  0.84714(7) & 0.98472(4) & 4.45086(3) \\			\hline
112&  0.84723(6) & 0.98437(4) & 4.45115(2)\\	 		\hline
119&  0.84734(5) & 0.98446(6) & 4.45344(4)\\			\hline
126&  0.84750(3) & 0.98448(4) & 4.45434(2)\\			\hline
133&  0.84759(3) & 0.98429(5) & 4.45567(2)\\			\hline
140 & 0.84772(4) & 0.98428(2) & 4.45648(2) \\	    	\hline
175 & 0.84782(5) & 0.98418(2) & 4.45902(2) \\	\hline	\hline      
$\infty$&0.8482(1)&0.9839(2)& 4.4641(3)\\ \hline  
		\end{tabular}
	\end{ruledtabular}
	\label{tab:2nn3d}
	\end{table}
	
In the second method, $\mu_c(L)$ is taken to be value of $\mu$ at which the susceptibility, $\chi$ is maximum. We then extrapolate to infinite $L$ using Eq.~(\ref{eq:muscaling}) (see Fig.~\ref{fig:mubylsq}). We obtain $\mu_c=4.4641(3)$. To determine the critical pressure, we first find  $ P $ at $\mu_c(L)$ for each $ L $. By extrapolating to infinite system size, we obtain the critical pressure to be $P_c= 0.6397(1) $.

\begin{figure}
	\includegraphics[width=\columnwidth]{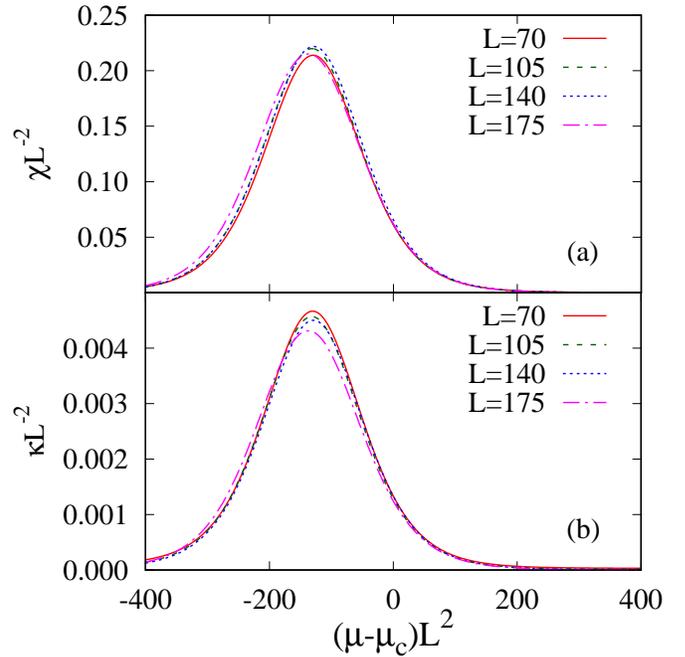}
	\caption{ The data for (a) susceptibility $\chi$ and (b) compressibility $\kappa$ for different $L$ collapse onto a single curve when scaled as in Eqs.~(\ref{eq:chiscaling}) and (\ref{eq:kappascaling}) with $\mu_c=4.4641$.  The data are obtained using SCWL algorithm.}
	\label{fig:collapse}
\end{figure}

The response functions $\chi$ and $\kappa$ obey the following finite size scaling relations for a first order phase transition, in two dimensions:
\bea
\chi &\approx& L^2 f_\chi [(\mu-\mu_c)L^2],
\label{eq:chiscaling}\\
\kappa &\approx& L^2 f_\kappa [(\mu-\mu_c)L^2],
\label{eq:kappascaling}
\eea
where  $f_\chi$ and $f_\kappa$ are scaling functions.  As shown in Fig.~\ref{fig:collapse}, the data for different $L$ collapse onto one curve when scaled as in Eqs.~(\ref{eq:chiscaling}) and (\ref{eq:kappascaling}), and hence consistent with a discontinuous transition. 

\section{\label{sec:summary}Summary and Discussion }

In this paper, we have studied in detail the phase diagram and the nature of the phase transition in the 3-NN lattice gas model on the triangular lattice. We have  shown that there is a single discontinuous transition from a disordered fluid phase to an ordered sublattice phase as density is increased from zero to one. The critical chemical potential,  coexistence densities and critical pressure are determined to be $\mu_c=4.4641(3)$, $\rho_f=0.8482(1)$,  $\rho_s=0.9839(2)$ and $P_c= 0.6397(1)$.

Our results are in contradiction with those of Ref.~\cite{darjani2019liquid}, where the existence of a hexatic phase with continuously varying exponents was claimed. The discrepancy is most likely due to issues with equilibration. In Ref.~\cite{darjani2019liquid}, the Monte Carlo simulations were done using either adsorption and local diffusion moves or using desorption and local diffusion moves. When the density is increased to the coexistence densities, the system should phase separate into fluid and solid phases. However, the phase separation usually occurs over time scales that diverge with system size. For hard core lattice gases, there is an additional issue of the system being trapped in long lived metastable states. Local diffusion moves alone make it difficult to equilibrate the system and can lead to identification of spurious phases, which we believe is the reason for a signature of a hexatic phase to be seen. We note that it is expected that for a much large range of exclusion than the third nearest neighbor, a hexatic phase should be present like in two dimensional discs, and the phase with true long-range order shrinks to zero thickness.

The critical chemical potential $\mu_c$ was determined using the tensor renormalization group method to be $\mu_c=4.4488$~\cite{akimenko2019tensor}. It matches up to the first decimal place with our estimate $\mu_c=4.4641(3)$. The tensor renormalization group method appears to be a good candidate,  different from SCUA and SCWL, for studying hard core lattice gases. It would be interesting to see which method gives more accurate results.  Another problem where they could be compared is to look at the results for models with larger exclusion. Preliminary results, using tensor renormalization group method, exist for $k=4$ and $k=5$~\cite{akimenko2019tensor}. Obtaining results for these models as well as for larger $k$ using the flat histogram algorithm used in this paper is a promising area for future study.

We also find that the flat histogram algorithm with cluster moves has certain advantages over the corresponding grand canonical algorithm beyond the fact the density of states allows thermodynamic quantities to be determined for all chemical potential. Beyond $L=70$, we find that the grand canonical SCUA finds it difficult to equilibrate the system in the coexistence region leading to quite large hysteresis loops when the data for increasing and decreasing parameters are compared. This issue is overcome in the flat histogram algorithm where we find that the algorithm is able to avoid issues of equilibration at high densities and is even able to access fully packed states quite easily. In the co-existence regime, the entropy is non-convex for finite system sizes.This feature leads to non-convexity in the entropy. We have exploited non-convexity to obtain accurate estimates for the coexistence densities. 

\begin{acknowledgments}
The simulations were carried out on the high performance computing machines Nandadevi at the Institute of Mathematical Sciences  and  the computational facilities provided by the University of Warwick Scientific Computing Research Technology Platform.
\end{acknowledgments}

\section*{Data Availability Statement}
The data that support the findings of this study are available from the corresponding author upon reasonable request.

\appendix
\section{ \label{appa} Strip cluster update algorithm (SCUA)}
Lattice gas model algorithms with simple evaporation and deposition has difficulty in equilibrating at higher densities. To overcome this, SCUA was proposed and were successful in equilibrating up to densities 0.99~\cite{2012-krds-aipcp-monte,2013-krds-pre-nematic,nath2014multiple,2014-kr-pre-phase,2015-rdd-prl-columnar,2018-pre-mnr-phase,vigneshwar2019phase}. 
In SCUA, a Monte Carlo move consists  following steps. Consider a row of $L$ sites in any diagonal along one of the three principal directions. Remove all the particles that are on this row. Some of the empty sites on this row are excluded from reoccupation due to presence of particles in nearby rows. 
The re-occupation of the row with a new configuration of particles, with the correct equilibrium probability in the grand canonical ensemble, can be done independently for each of the empty intervals. Thus, the problem of re-occupation reduces to finding the probability of different configurations in a one-dimensional lattice of length $\ell=1,2,\ldots L$.

 Let $\Omega_o(\ell)$ denote the partition function on a one dimensional lattice of $\ell$ sites 
with open boundary conditions. It obeys the recursion relation $\Omega_o(\ell) = z \Omega_o(\ell-3) + \Omega_o(\ell-1)$, for
$\ell\geq 3$, with the boundary conditions $\Omega_o(0)=1$, $\Omega_o(1)=1+z$, and $\Omega_o(2)=1+2 z$. The 
probability that the left most site is occupied by a  particle is $p(\ell)=z \Omega_o(\ell-3)/\Omega_o(\ell)$. 
If left empty (with probability $1-p(\ell)$), we consider 
the neighbor to the right and reduce the number of lattice sites by one. 
If the first site is occupied by a  particle (with probability $p(\ell)$), then we skip two more lattice sites and reduce the number 
of lattice sites by $3$, and repeat the procedure. The partition functions $\Omega_o(\ell)$ diverge exponentially with $\ell$. Hence, it is computationally easier to express the recursion relations in terms of the probabilities. From the recursion relations for the 
partition functions, it is straightforward to show that the probabilities $p(\ell)$ satisfy the recursion relation 
\be
p(\ell) = \frac{p(\ell-1)}{1+p(\ell-1)-p(\ell-3)},~~\ell=3, 4, \ldots,
\ee
with the boundary conditions $p(0)=0$, $p(1)= z/(1+z)$, and $p(2) = z/(1+2 z)$.
\begin{figure}
	\includegraphics[width=\columnwidth]{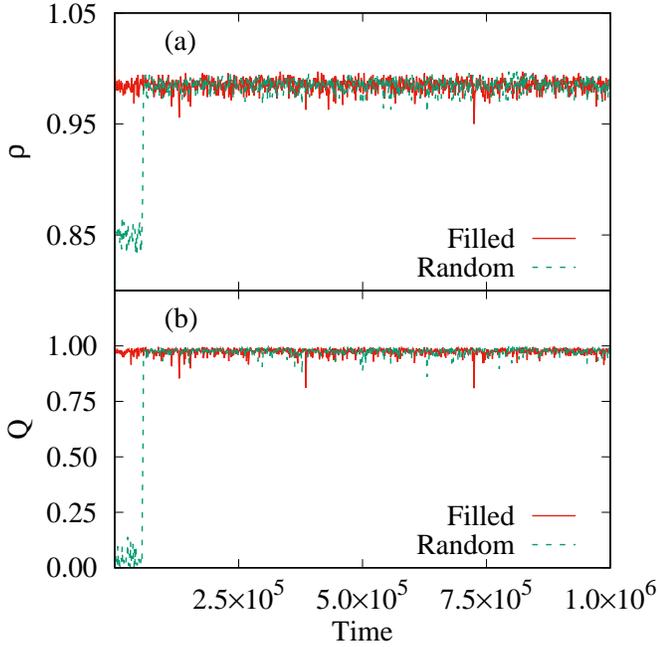}
	\caption{\label{fig:equilibration} The time evolution of (a) density, $\rho$ and (b) order parameter, $Q$ for two different initial conditions, one which has sublattice order (labeled as Filled) and other which is obtained from random deposition (labeled as Random). The data are for $\mu=4.50$ and $L=70$.  }
\end{figure}

For periodic boundary conditions (when $\ell=L$), the recursion relations have to be modified. Let 
$\Omega_p(\ell)$ be the partition function of a one dimensional lattice of 
length $\ell$ with periodic boundary conditions. It is easy to see that 
$\Omega_p(\ell) =  2 z \Omega_o(\ell-5) + \Omega_o(\ell-2)$. Consider 
$p_{pbc}(\ell)= 2 z \Omega_o(\ell-3) / \Omega_p(\ell)$. $p_{pbc}(\ell)$ should be identified as the probability that either the first or second site is occupied by a particle. Using the recursion relations, it is straightforward to derive
\be
p_{pbc}(\ell)=  \frac{2 p(\ell-2)}{1+2p(\ell-2)}.
\ee
The relevant probabilities are stored in a  list to reduce the computation 
time. We repeat the evaporation and re-occupation of a row by  particles for each of the $3L^2$ rows. It is straightforward to see that the algorithm is ergodic, 
and satisfies the detailed balance condition. Also, the updating of rows that are separated by five rows are independent of each other and can be performed simultaneously.

We first show that, using the algorithm, we are able to equilibrate the 3-NN model for densities as large as $0.99$. In Fig.~\ref{fig:equilibration}, we show the temporal evolution for density, starting from two different initial conditions: one which is fully packed and the other at a lower density generated by depositing particles at random. It is clear that the system equilibrates at a density $\approx 0.99$, independent of the initial condition [see Fig.~\ref{fig:equilibration} (a)]. In Fig.~\ref{fig:equilibration} (b), we show the corresponding time evolution for the order parameter [see Eq.~(\ref{eqn:op}) for definition]. The two initial conditions correspond to the order parameter being initially $1$ and zero. It can be seen that the system equilibrates to the same value of order parameter at large times.

\section{\label{appb} Benchmarking SCWL algorithm with SCUA.}

\begin{figure}
\includegraphics[width=\columnwidth]{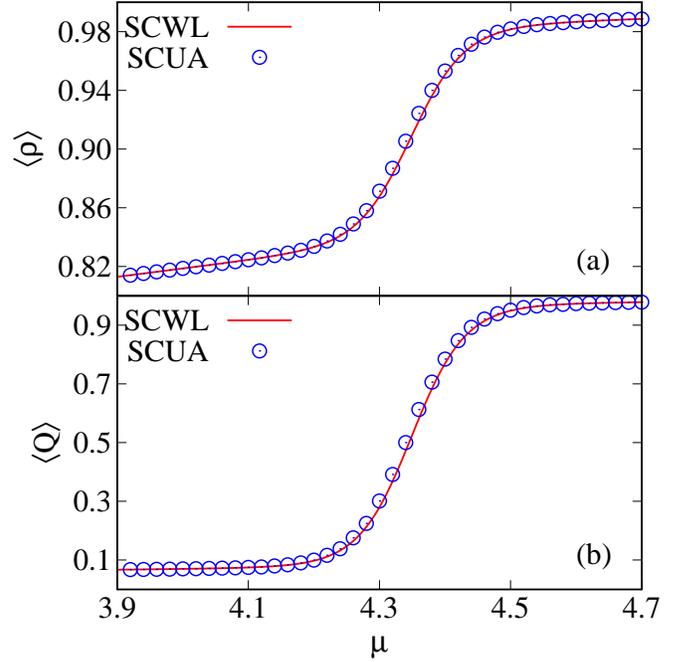}
\caption{\label{fig:compare_mc_wl}Variation of (a) density, $\rho$, and (b) order parameter $Q$ with $\mu$. SCWL and SCUA refer to the data obtained from the strip cluster Wang Landau and  the strip cluster update (grand canonical) algorithm respectively. The data are for  $L=35$. }
\end{figure}

To benchmark  the  flat histogram SCWL algorithm, we compare the results obtained from the method with those from  the grand canonical Monte Carlo simulations at fixed $\mu$ using SCUA. The data for order parameter and density for $L=35$ are shown in Fig.~\ref{fig:compare_mc_wl}.  The results obtained from both simulations show good agreement.

Note that in the coexistence regime there appears to be a systematic deviation in the results from the two simulations. In the grand canonical SCWL simulations, the results in the co-existence regime depend on how many  transitions from the low density phase to high density phase are averaged over. When infrequent transitions are there, systematic errors can set in. Likewise, in the SCUA simulations, there is the possibility of the system not fully phase separating in the co-existence region. The slight discrepancy in data is possibly due to the above reasons.
\section{\label{appc} Canonical simulations at fixed density}

In this appendix, the algorithm for the  fixed-density  Monte Carlo simulations is described. First, we deposit particles till the desired density (in this case $0.920$) is reached. As the desired density is high, random deposition of particles will not be useful to reach that density. The particles are initially deposited in only one sublattice. Given any valid configuration, the system evolves as follows. A site is chosen at random. If occupied, then the particle is removed and  placed at another randomly chosen site provided the hard core constraint is not violated. If disallowed, the particle is placed on the original site.  One Monte Carlo step consists of $L^2$ such moves. We have used $10^6$ Monte Carlo steps to equilibrate the system of $L=140$. The results obtained from the simulation is shown in Fig.~\ref{fig:snap_canonical}, where we see a clear phase separation of the disordered phase and the sublattice phase at $\rho=0.920$.

%

\end{document}